\pdfoutput=1
\documentclass[sigconf]{acmart}

\AtBeginDocument{%
  \providecommand\BibTeX{{%
    \normalfont B\kern-0.5em{\scshape i\kern-0.25em b}\kern-0.8em\TeX}}}

\setcopyright{acmlicensed}
\copyrightyear{2024} 
\acmYear{2024} 
\setcopyright{acmlicensed}\acmConference[ICSE '24]{2024 IEEE/ACM 46th
International Conference on Software Engineering}{April 14--20, 2024}{Lisbon,
Portugal}
\acmBooktitle{2024 IEEE/ACM 46th International Conference on Software
Engineering (ICSE '24), April 14--20, 2024, Lisbon, Portugal}
\acmDOI{10.1145/3597503.3639230}
\acmISBN{979-8-4007-0217-4/24/04}

\usepackage{amsmath,amsfonts}
\usepackage{multirow}
\usepackage{algorithmic}
\usepackage{graphicx}
\usepackage{textcomp}
\usepackage{color}
\usepackage{xcolor}
\usepackage{framed}
\usepackage{listings}
\usepackage{bbding} 
\usepackage{soul}
\usepackage{enumitem}
\usepackage{threeparttable}
\usepackage{amsmath, bm}
\usepackage{algorithm}
\usepackage{pifont}
\usepackage[skins]{tcolorbox} 
\usepackage[normalem]{ulem}
\tcbuselibrary{breakable} 
\usepackage[ruled,lined,linesnumbered,vlined,algo2e]{algorithm2e}
\def\BibTeX{{\rm B\kern-.05em{\sc i\kern-.025em b}\kern-.08em
    T\kern-.1667em\lower.7ex\hbox{E}\kern-.125emX}}
\begin{document}

\title{Empirical Analysis of Vulnerabilities Life Cycle in Golang Ecosystem}

\newcommand{\mynote}[2]{
    \fbox{\bfseries\sffamily\scriptsize#1}
    {\small$\blacktriangleright${\emph{#2}}$\blacktriangleleft$}}
\newcommand{\rev}[1][\textcolor{magenta}]{#1}
\newcommand{\ly}[1][\textcolor{magenta}]{#1}%
\newcommand{\chengwei}[1][\textcolor{black}]{#1}
\newcommand{\cw}[1]{\textcolor{violet}{(CW:#1)}}
\newcommand{\jc}[1]{\textcolor{black}{#1}}
\newcommand{\jcrevision}[1]{\textcolor{black}{#1}}
\newcommand{\short}[1]{\textcolor{red}{#1}}
\renewcommand{\shortauthors}{Hu, et al.}


\author{Jinchang Hu}
\orcid{0000-0002-1483-8854}
\authornote{These authors contributed equally to this work}
\affiliation{%
  \institution{College of Command and Control Engineering, Army Engineering University of PLA}
    \city{Nanjing}
  \country{China}
}
\email{hujinchang@aeu.edu.cn}

\author{Lyuye Zhang}
\orcid{0000-0003-3087-9645}
\authornotemark[1]
\affiliation{%
  \institution{Continental-NTU Corporate Lab, Nanyang Technological University}
    \city{Singapore}
  \country{Singapore}
}
\email{zh0004ye@e.ntu.edu.sg}

\author{Chengwei Liu}
\orcid{0000-0003-1175-2753}
\authornote{These authors are the corresponding authors.}
\affiliation{%
  \institution{School of Computer Science and Engineering, Nanyang Technological University}
    \city{Singapore}
  \country{Singapore}
}
\email{chengwei001@e.ntu.edu.sg}

\author{Sen Yang}
\affiliation{%
  \institution{Academy of Military Science}
    \city{Nanjing}
  \country{China}
}
\email{yangsen0310@aeu.edu.cn}

\author{Song Huang}
\authornotemark[2]
\affiliation{%
  \institution{College of Command and Control Engineering, Army Engineering University of PLA}
    \city{Nanjing}
  \country{China}
}
\email{huangsong@aeu.edu.cn}

\author{Yang Liu}
\orcid{0000-0001-7300-9215}
\affiliation{%
  \institution{School of Computer Science and Engineering, Nanyang Technological University}
  \city{Singapore}
  \country{Singapore}}
\email{yangliu@ntu.edu.sg}


\begin{abstract}
{Open-source software (OSS) greatly facilitates program development for developers. However, the high number of vulnerabilities in open-source software is a major concern, including in Golang, a relatively new programming language. 
In contrast to other commonly used OSS package managers, Golang presents a distinctive feature whereby commits are prevalently used as dependency versions prior to their integration into official releases. This attribute can prove advantageous to users, as patch commits can be implemented in a timely manner before the releases. However, Golang employs a decentralized mechanism for managing dependencies, whereby dependencies are upheld and distributed in separate repositories. This approach can result in delays in the dissemination of patches and unresolved vulnerabilities.
}

{To tackle the aforementioned concern, a comprehensive investigation was undertaken to examine the life cycle of vulnerability in Golang, commencing from its introduction and culminating with its rectification. 
To this end, a framework was established by gathering data from diverse sources and systematically amalgamating them with an algorithm to compute the lags in vulnerability patching. It turned out that $66.10\%$ of modules in the Golang ecosystem were affected by vulnerabilities.
\chengwei{Within the vulnerability life cycle, we found two kinds of lag impeding the propagation of vulnerability fixing.}
By analyzing reasons behind non-lagged and lagged vulnerabilities, 
timely releasing and indexing patch versions could significantly enhance ecosystem security.
}



\end{abstract}




\maketitle

\section{Introduction}\label{sec:intro}


The growth of software complexity has boosted the adoption of third-party libraries (TPLs) as dependencies, to reduce development costs. However, this also poses a new threat that vulnerabilities in TPLs can be widely propagated through dependencies.
For instance, CVE-2022-41723~\cite{CVERecor90:online} existing in \textit{golang.org/x/net}, a fundamental Golang networking library, could influence over 170K downstream packages, according to Open Source Insights~\cite{GO20231557:online}.

Many existing works and tools have been proposed to demystify the impact of TPL vulnerabilities in different ecosystems. Some researchers~\cite{Alfadel2021,Liu2022a,Prana2021,Shahzad2020} investigate the vulnerability impact and their life spans from the perspective of ecosystems, some researchers~\cite{gonzalez2020characterizing,Decan2018,Imtiaz2022,Stringer2020,Zerouali2019} conduct empirical studies on technical lags from the perspective of user projects (i.e., the delays of upgrading vulnerable dependencies). Tools like SCA (Software Composition Analysis)~\cite{sca} are also proposed to identify TPL vulnerabilities in user projects.
Moreover, some existing work~\cite{cogo2023understanding, alfadel2021use, Zhang2022} study the adoption of Dependabot~\cite{dependabot} to remediate vulnerable dependencies.
All these existing works are conducted in mature ecosystems that are established on package managers, supported by centralized registries following Semantic Versioning~\cite{semver} (SemVer).

However, as the first attempt at decentralized registries, Golang embraces the Git system to manage dependencies~\cite{AboutGit94:online}. 
Specifically, unlike traditional centralized registries where developers mostly define dependencies by released version tags, \chengwei{commits are widely used as version references in Golang.} 
Though Golang Index~\cite{golangindex}, a centralized proxy, was introduced to mirror packages, there are still huge differences in broadcasting newly released versions (e.g., patch versions), between Golang and other mature ecosystems. 
For instance, patch versions could be available to all users, or even automatically integrated, during new installations (i.e., NPM dependency tree changes~\cite{Liu2022a}).
However, in the Golang ecosystem, only versions explicitly utilized by users can be cached by Golang Index according to the documentation~\cite{GoModule18:online}.
Hence, methodologies concluded from traditional ecosystems could be compromised in Golang, and none of the existing work has yet researched the vulnerability life cycle under the special Golang dependency mechanism.

\jc{In this paper,}
we aim to demystify the existence of vulnerabilities and the delays of vulnerability fixes so that solutions can be derived, to mitigate ecosystem-wide vulnerability impact in the Golang ecosystem. To fill these gaps, we face the following challenges: (1) \textbf{Data Scope.} As Golang allows versions of dependencies to be defined as commits, it is unrealistic to record all commits of all Golang libraries as potential versions. Moreover, Golang Index only indexes libraries that are imported as dependencies by downstream dependents via the Golang official tool, Go Modules~\cite{GoModule18:online}, thus it is tricky to ensure completeness when collecting existing Golang packages. (2) \textbf{Vulnerability Mappings.} 
Vulnerabilities in Golang libraries are not easily mapped to specific version ranges due to the usage of commits as dependency versions, making version-level mappings inapplicable.
Hence, finer-grained vulnerability mappings are necessary when analyzing vulnerability impact in Golang, while such commit-level vulnerability mapping is barely captured by existing advisories. (3) \textbf{Vulnerability Life Cycle.} Golang has its unique distributed mechanism for releasing new packages, in this case, the life cycle and spread of vulnerabilities and corresponding patches across the ecosystem, are different from other ecosystems\jc{,} and it is yet to be systematically investigated.

Towards the challenges, we first constructed an infrastructure by collecting data from multiple sources\jc{,} and uniformly integrating them into an analytical framework. The framework maps the foundational vulnerabilities to modules, commits from repositories, and vulnerable dependents. Then we implemented an algorithm to identify critical time points in the life cycles of vulnerabilities\jc{,} and calculate the lags of fixing vulnerabilities for the study. Specifically, for challenge (1), to accurately locate the commit-level vulnerable ranges of vulnerabilities, we scrutinized the reference links from NVD~\cite{NVDHome98:online} and Snyk~\cite{SnykDeve34:online} databases to identify the fixing commits. For challenge (2), to avoid analyzing all commits, beginning with identified fixing commits, we backward-tracked the vulnerable commits and forward-tracked the patched commits. For challenge (3), besides the Golang Index~\cite{golangindex}, we collected the dependency relationships from the Open Source Insight~\cite{osinsightdata} and verified them with Golang configuration files from the respective repositories.

\begin{figure*}[!htbp]
  \centering
  \includegraphics[width=1\linewidth]{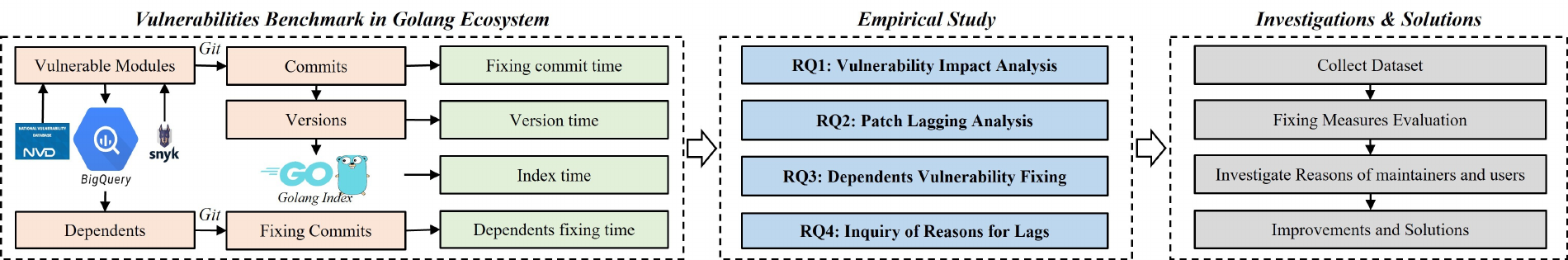}
  \caption{Overview of our work}
  \label{fig:overview}
\end{figure*}

To this end, 
we first identified fixing commits associated with vulnerabilities in the vulnerable modules.
This facilitated the comprehensive understanding of vulnerabilities and the analysis of the fixing time lag. Then, with the measured time lags, we determined the impact of the lags by evaluating the fixing tendency by dependents. Finally, to provide solutions for mitigating the vulnerability in the ecosystem, we inquired the community about the reasons for the patch lagging. Furthermore, we aim to answer four research questions to evaluate the effectiveness of Golang's dependency management mechanism\jc{,} for reducing the vulnerability life cycle:
\begin{itemize}[leftmargin=*]
\item {\bfseries{RQ1:} Vulnerability Impact Analysis.} To what extent could TPL vulnerabilities affect modules in the Golang ecosystem?
\item {\bfseries{RQ2:} Patch Lagging Analysis.} How long does it take to release and index the patch versions?
\item {\bfseries{RQ3:} Dependents Vulnerability Fixing.} What are the fixing lags by dependents, and quantitatively what factors could facilitate the fixing?
\item { \bfseries{RQ4:} Inquiry of Reasons for Lags.} What are the reasons for the patch lagging regarding both module maintainers and users?
\end{itemize}

Through our study, we have quantitatively substantiated that vulnerabilities have been increasingly affecting the Golang ecosystem, with $66\%$ of modules impacted by May 30, 2023. \chengwei{We discovered that delays in patch version release and indexing contributed to the persistence of these vulnerabilities, accounting for $10\%$ and $21\%$ of the vulnerabilities respectively.}
In contrast, our findings revealed that both timely patch version release and indexing can effectively lessen the influence of vulnerabilities, as gauged by the number of dependents that explicitly address vulnerabilities.
Beyond those developers who actively rectify vulnerabilities, we sought to assist the community in enhancing vulnerability remediation. To this end, we questioned maintainers and users regarding the underlying causes of bad practices such as delayed release and indexing. Drawing on their responses, we have summarized a set of recommendations for the Golang community to bolster vulnerability mitigation.

In summary, this paper makes the following contributions:
\begin{itemize}[leftmargin=*]
\item We quantitatively substantiated that the vulnerabilities have significantly impacted the Golang ecosystem. 
\item We conducted a large-scale study to evaluate the impact of time lags on vulnerability life cycles in the Golang ecosystem.
\item We interacted with maintainers and users 
to gain insights into the reasons for delayed patching, and offered recommendations to potentially shorten the vulnerability life cycle. All data, scripts, and queries used in the paper is available at our website~\cite{dataset}.
\end{itemize}

\section{Background and Motivating Example}

\subsection{Background}

Key concepts used throughout this paper are explained:\\
{\bfseries{1) Module}}: A collection of packages ~\cite{GoModule22:online} that are released and versioned together, similar to the concept \textit{library} in other ecosystems. \\
{\bfseries{2) Version}}: A version identifies an immutable snapshot of a module.
A version tagged by the maintainer usually starts with the letter $v$, followed by a semantic version~\cite{semver}.\\
{\bfseries{3) Pseudo-version}}: A pseudo-version~\cite{GoModule83:online} \footnote{For example, v0.0.0-20191109021931-daa7c04131f5 is a pseudo-version.} is a uniformly formatted pre-release version based on a commit. A pseudo-version comprises three parts, a base version prefix (\textit{vX.Y.Z-0}), which is either derived from a semantic version that precedes the revision, a UTC timestamp of the commit time, and a commit hash.\\
{\bfseries{4) Index}}: When a module is declared as a dependency by any user, it is registered in the Golang Index. Using \texttt{module@latest} retrieves the latest version from the Golang Index. When no version is available, a pseudo-version is retrieved. \\
{\bfseries{5) Fixing Commit}}: A commit involving the code changes that address a vulnerability. The commit could be incorporated into the subsequent version releases. Note that, in a repository, the fixing code changes could be accommodated into multiple commits for multiple pipelines of releases which forms a complex commit graph.\\
\chengwei{{\bfseries{6) Patch Version}}: It refers to the first subsequent stable non-pre-release version that incorporates the fixing commit following the vulnerable versions.}\\
{\bfseries{7) File go.mod $\&$ go.sum}}: 
\textit{go.mod} is a Golang project's dependency declaration file, listing direct and indirect modules with versions. \textit{go.sum} is automatically generated during project building, /recording dependencies and checksums. 

\subsection{Motivating Example}

Since versions could only be automatically pushed to users when they are officially indexed by Golang, the time interval between the release and indexing could cause unnecessary lag. If the new version includes patches for vulnerabilities, the lag could result in a window period for attackers. 
For example, a widely-used module \emph{go.etcd.io/etcd} (\textit{etcd}) owning $43.0k$ stars on GitHub, was known to have a vulnerability, \emph{CVE-2020-15113}. The patched version, \emph{v3.3.23}, was released on July 17, 2020, in response to this vulnerability. For downstream users, obtaining the patch version \emph{v3.3.23} through the Golang client is not possible if the version has not been indexed by Golang beforehand. 

Due to the absence of a real-time registry for Golang, the explicit specification requires developers to manually search \emph{v3.3.23} in the \textit{etcd} repository.
Worse, since \textit{v3.3.23} version was quickly superseded by \textit{v3.3.24} on August 19, 2020,\textit{v3.3.23} was not even used by any dependent.
After indexing patch versions after one month, $109$ dependents performed the fixing by either upgrading \textit{etcd} or migrating to other modules. \chengwei{While such situations cannot be mitigated by modern SCA tools because they mostly fail to handle Golang-specific mechanisms. Taking Dependabot as an example, it does not recommend pseudo-versions or un-indexed patch versions according to the documentation~\cite{dependabot_update}, source code~\cite{dependab41:online,dependab81:online}. }
To understand to what extent the lag affects vulnerability fixing and how to mitigate the lag, we conducted this study.

\section{METHODOLOGY}
We first constructed an infrastructure for data analysis given multiple data sources. Then, we derived four critical periods and two types of lag in the life cycle of a vulnerability with algorithm~\ref{algorithm:Lags}.

\subsection{Analytical Infrastructure Construction}
As illustrated in Figure~\ref{fig:overview}, the left part demonstrates the structure of our analytical infrastructure. This research investigates the life cycles of vulnerabilities, starting with the acquisition of vulnerability data. Relevant references from public vulnerability databases were examined to identify the fixing commits in the repositories of the affected modules. Also, the subsequent patched versions were obtained from the commit history. The fixing commits and versions were used to initiate the lag analysis process. Additionally, the indexing time of versions was collected, given the unique indexing mechanism of Golang. To gain insight into the fixing operations of downstream dependents, the commit history of their repositories was further analyzed to obtain the dependent fixing time.


     \noindent $\bullet$ {\bfseries Vulnerability.} We collected \chengwei{$1,837$} Golang vulnerabilities from two databases: Snyk Advisory~\cite{SnykDeve34:online} and NVD~\cite{NVDHome98:online}. To locate the fixing commits in the vulnerable modules, we manually scrutinized the reference links and successfully derived \chengwei{$1,269$} vulnerabilities with fixing commits, involving \chengwei{$441$} unique modules. 
     
     \noindent $\bullet$ {\bfseries Dependency Relation.} Based on the vulnerable modules, we acquired the dependents that directly and indirectly use vulnerable modules as dependencies from Open Source Insight~\cite{osinsightdata}. For each dependency relationship, the module name and module versions of both dependencies and dependents were kept. The study identified dependent module names belonging to vulnerable modules.
     
     \noindent $\bullet$ {\bfseries Indexing time of version.} The Golang Index ~\cite{GoModule18:online} registers the timestamp and dependency relationship of a newly released version when the version is used by dependents for the first time. These timestamps refer to the moment when the new versions are publicly known to Golang developers. The crawled records include timestamps, module names, and versions. However, versions that are not indexed were not recorded.
    
     \noindent $\bullet$ {\bfseries Commit history.} We collected the commit history for both vulnerable modules and dependents, including the commit ids, commit subjects and messages, associated versions, commit time, and the relationships among commits. Atop the commit history, the fixing lag analysis will be conducted.

\subsection{Fixing Lag Analysis}
\label{sec:alg}

\begin{figure*}[!htbp]
  \centering
  \includegraphics[width=0.9\linewidth]{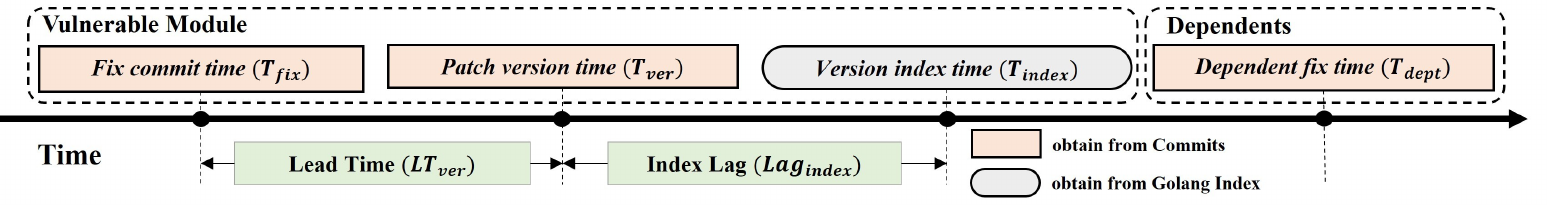}
  \caption{\chengwei{The fixing lags of Golang vulnerabilities in their lifecycles.}}
  \label{fig:Technical lag}
\end{figure*}

\chengwei{In this subsection, we aim to precisely identify the time lag of vulnerability fixing. 
As illustrated in Figure~\ref{fig:Technical lag}, there are three major timestamps, 
fixing commit time $T_{fix}$, version release time $T_{ver}$, and indexing time $T_{index}$. $T_{fix}$ refers to the time when fixing commits are merged or committed to the main branch. For a vulnerable module, the duration between $T_{fix}$ and $T_{ver}$ (following version release time), is denoted as the Lead Time $LT_{ver}$. Although $LT_{ver}$ usually accommodates the time for code review and testing, it would still cause risky lags for downstream libraries if it is unusually long, we name such lags as version lag, $Lag_{ver}$. 
Moreover, we denote the first time of patch versions indexed in Golang Index as the indexing time ($T_{index}$),
and the time from $T_{ver}$ to $T_{index}$ is denoted as the index lag $Lag_{index}$. From the dependent side, 
the time when the vulnerable dependency is addressed is denoted as $T_{dept}$.}

\subsubsection{\bfseries Lead Time}
\chengwei{\emph{Lead Time} is the time interval between $T_{fix}$ and $T_{ver}$. To obtain $T_{fix}$ and $T_{ver}$, we first downloaded repositories of vulnerable modules and used git commands to obtain the commit trees. Then, For each vulnerability, from reference links, fixing commits are obtained. Because the code changes that fix the commits could be accommodated in other commits, we searched for similar commits from the vulnerable module's commit tree based on the subjects of fixing commits in Alg~\ref{algorithm:Lags} L2-L4. Among these commits, the time of the earliest fixing commit is considered as $T_{fix}$. As fixing commits could be committed on non-release branches for development instead of direct use, we further searched the subsequent merging commit along the commit chain. If a merging commit is spotted, $T_{fix}$ is updated to the time of it as in L8.
For each fixing commit, we matched the commit id with the versions' commit ids to locate the timestamps of the first subsequent version after the fixing commit as in L9-L14. The timestamp of the first version is denoted as $T_{ver}$. Finally, we calculated the $LT_{ver}$ for each vulnerability.}


\subsubsection{\bfseries Version Lag}
\chengwei{Since unusually long $LT_{ver}$ could easily postpone the tagging of patch versions, we further empirically measured such \emph{version lag}, i.e., $Lag_{ver}$, based on $LT_{ver}$. We narrow down vulnerabilities by two prerequisites: \ding{172} $LT_{ver}$ > 1 week; \ding{173} current release cycle is greater than the normal version release cycle. For these vulnerabilities, $Lag_{ver}$ is defined as the exceeding time of the current cycle over the normal release cycle in L23. 
Specifically, the process retrieves the time interval between each pair of adjacent tags on the commit tree. Based on this, The quartile formula~\cite{Quartile6:online} is then utilized to ascertain the upper bound of the normal length of version release cycle ranges by $Q3 + 1.5 * (Q3 - Q1)$, as the filter of calculating $Lag_{ver}$.
}



\subsubsection{\bfseries Index Lag}
\emph{Index Lag} refers to the time interval between the earliest version release time $T_{ver}$ and the earliest index time of versions in Golang Index $T_{index}$. 
Besides $T_{ver}$, we determined $T_{index}$ by searching for the index time of versions at Golang Index~\cite{httpsind37:online} by module names and versions. The earliest timestamp is denoted as $T_{index}$. Then, $Lag_{index}$ could be calculated as in L25.

\subsubsection{\bfseries Dependent Fix Time}
\chengwei{\emph{Dependent Fix Time} ($T_{dept}$) is the time of the first dependent that fixes the vulnerability in its dependencies.} To measure $T_{dept}$, we first identified all the dependents that used the vulnerable versions of vulnerable modules.
\jc{For each dependent, we retrieved the git commit history of the \textit{go.sum} lock file, which specifies dependencies and their versions. Our goal was to locate the commit where the vulnerable version was excluded (Algorithm~\ref{algorithm:Lags} L26-L32). To identify the fixing commit for vulnerabilities within a project's dependency tree, we traversed the modification history of \textit{go.sum} for each dependent. Starting from the latest commit that altered the \textit{go.sum}, we exhaustively searched for the existence of the vulnerable version. If not found, we move to the previous commit that altered \textit{go.sum}, continuing until no such commit was found. Once the vulnerable version was located, the following commit in which the vulnerable versions are excluded was considered as the dependent's fixing commit, and its timestamp was denote as the $T_{dept}$ for the dependent. 
\chengwei{Note that the fix commit may have removed the vulnerable dependency or upgraded the vulnerable versions to clean versions.} Based on Algorithm~\ref{algorithm:Lags}, we can calculate the two types of lags, $Lag_{ver}$ and $Lag_{index}$, for vulnerable dependents, as well as $T_{dept}$ for each dependent.}


\makeatletter
\newenvironment{breakablealgorithm}
  {
   \begin{center}
     \refstepcounter{algorithm}
     \hrule height.8pt depth0pt \kern2pt
     \renewcommand{\caption}[2][\relax]{
       {\raggedright\textbf{\ALG@name~\thealgorithm} ##2\par}%
       \ifx\relax##1\relax 
         \addcontentsline{loa}{algorithm}{\protect\numberline{\thealgorithm}##2}%
       \else 
         \addcontentsline{loa}{algorithm}{\protect\numberline{\thealgorithm}##1}%
       \fi
       \kern2pt\hrule\kern2pt
     }
  }{
     \kern2pt\hrule\relax
   \end{center}
  }
\makeatother

\begin{algorithm2e}[t!]
    \small
 \setcounter{AlgoLine}{0}
 \caption{Calculation of Lags}
 \label{algorithm:Lags}
 \DontPrintSemicolon
 \SetCommentSty{mycommfont}
 {
 \KwIn{$dep\_commits$: commits of a vulnerable dependency,  $dept\_commits$: commits of a dependent}
     \KwOut{\emph{Tag\_Lag}s, \emph{Index\_Lag}s}
        $fix\_commit \gets fromNVD(vul)$\;
        
        \ForEach {$commit \in dep\_commits$}{
            \If{$commit.subject == fix\_commits.subject$}{   
                $fix\_commits.add(commit)$\;
            }
        }
        $commit\_stack \gets fix\_commits$\;
        \chengwei{$T_{fix} = MIN(fix\_commits.release\_time)$\;
        \If{$subsequenct(T_{fix})$ is $merge$}{
            $T_{fix} = subsequenct(T_{fix})$\;
        }
        }
        $commit \gets commit\_stack.pop$\;
        \While {$commit \ne \phi$}{
            $non\_vul\_commits \gets non\_vul\_commits + commit$\;
            \ForEach {$son\_commit \in commit.son\_commits$}{
                $ commit\_stack \gets commit\_stack + son\_commit$\;
            }   
            $commit \gets commit\_stack.pop$\;
        }
        \ForEach {$commit \in non\_vul\_commits$}{
            \If{$isVersion(commit)$ $is$ $True$}{        
                $non\_vul\_versions.add(commit$)\;
            }
        }
        \chengwei{
        $normalCycle = MAX(quartiles(tagIntervals))$\;
        }
        $T_{ver} = MIN(non\_vul\_versions.release\_time)$\;
        \chengwei{
        $currentCycle = T_{previous\_ver} - T_{ver}$\;
        }

        \chengwei{$LT_{ver} = T_{ver} - T_{fix}$\;
        \If{$LT_{ver}>1week \&\& currentCycle>normalCycle$}{
            $Lag_{ver} = currentCycle-normalCycle$\;
        }
        $T_{index} = fromIndex(non\_vul\_version)$\;
        $Lag_{index} = T_{index} - T_{ver}$\;
        $com\_stack' \gets depts\_commits$\;
        \While{\text{$com\_stack' \ne \phi$}}{
            $com \gets com\_stack'.pop$\;
            \If {! $hasVulCommit(vul,com)$ and $hasVulCommit(vul,com_{prev})$}{
                $T_{dept} = com.release\_time$\;
                \bfseries{Break}\;
            }
            $com_{prev} = com$\;
        }
        }
        \KwRet {\text{$\chengwei{LT_{ver}, Lag_{ver}}, Lag_{index}, T_{dept}$}}
}
\end{algorithm2e}

\section{Empirical Study}

Considering the distinctive package management of Golang (i.e., decentralized registries and Golang index), the propagation of vulnerability patches could be lagged. In this case, we carry out an ecosystem-wide empirical study to investigate the vulnerability impact and the propagation of their corresponding patches by answering the research questions listed in Section~\ref{sec:intro}. 

\chengwei{Specifically, RQ1 assesses the existence of vulnerable dependencies in the Golang ecosystem. Subsequently, we investigated the reactions of maintainers and downstream dependents of vulnerable packages in RQ2 and RQ3. In RQ4, 
, we employed manual investigation and maintainer inquiries to unveil the reasons of the various lags that hinders vulnerability fixes.
}

\subsection{RQ1: Vulnerability Impact Analysis}
\label{sec:rq1}
\subsubsection{\textbf{Data Preparation}}
\label{sec:prepare}
We first collected $1,837$ Golang vulnerabilities from two mainstream databases, i.e., Snyk Advisory~\cite{SnykDeve34:online} and NVD~\cite{NVDHome98:online}. 
\chengwei{To substantiate the representativeness of our vulnerability dataset (1,837), we further compare the vulnerabilities with OSV~\cite{osv} (1,659) and Github Advisory~\cite{githubadvisory} (1,251) by OCT 15, 2023, the exceeded numbers of Golang vulnerabilities proves the representativeness of our dataset.}
\chengwei{Subsequently, we utilized the dataset from Open Source Insight~\cite{osinsightdata}, which maintained package and dependency data from various sources, to retrieve modules and their dependency relations of the Golang ecosystem. 
Given the absence of centralized registry in Golang, only modules that are imported as dependencies would be indexed, we are only able to take them as the entire scope of the Golang ecosystem. 
As a result, $725,286$ modules and $7,892,152$ versions are captured till May 2023, and $475,638,176$ dependency relations associated with the 1,837 vulnerabilities are identified. Moreover, Comparing to other mainstream platform (i.e., \textit{libraries.io}~\cite{librariesio}, $472K$ Golang modules by OCT 15, 2023), our dataset ($725k$) is also more complete and representative.}
For indexing time, we proceeded to retrieve the timestamps and dependency relationships of all versions from the Golang Index~\cite{GoModule18:online}. These timestamps represented the instances when the new versions became publicly known to the ecosystem.   


\subsubsection{\textbf{Impact of Golang Vulnerabilities}}
The $1,837$ vulnerabilities 
pose a notable impact on the Golang ecosystem that they
affect $479,411$ modules ($66.10\%$ of all modules) and $6,313,404$ versions ($80.00\%$ of all versions). 
By the data collection date, $455,813$ modules and $6,071,096$ versions have still not fixed vulnerabilities, which account for $62.85\%$ and $76.93\%$ of all in the Golang ecosystem.

\begin{figure}[]
  \centering
  \includegraphics[width=0.9\linewidth]{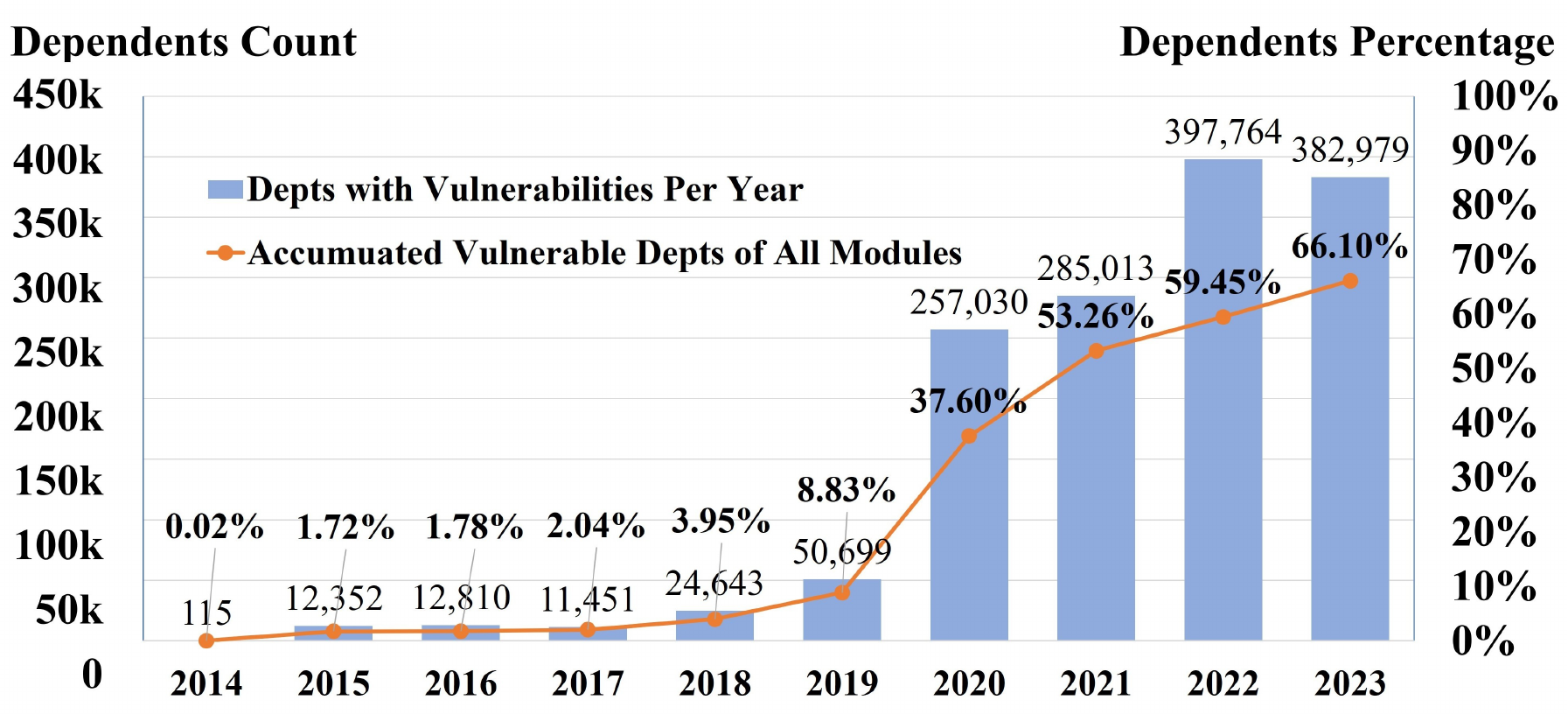}
  \caption{Distribution of Affected Depts}
  \label{fig:exist_vul}
\end{figure}

From the time perspective, the proportion of affected downstream dependents increases over the years as well. \chengwei{In Figure~\ref{fig:exist_vul}, the histogram plots the number of dependents still affected by vulnerabilities identified in each year, while the broken line plots the percentage of dependents which had been affected by vulnerabilities in each year. As can be observed, despite a notable increase in the number of dependents of vulnerabilities from 2020, it is anticipated that this trend will continue in 2023, given that the data was collected on May 30th.} This trend is supported by the rapidly growing proportion of accumulated dependents.

\begin{tcolorbox}[size=title,opacityfill=0.1,breakable,boxsep=1mm]
{\bfseries{Finding-1:}} 
The vulnerabilities affected a significant number of downstream dependents ($479,411$ $66.10\%$) by the data collection date.
$62.85\%$ of the dependents have still not fixed the vulnerabilities.
\end{tcolorbox}

\begin{figure}[]
  \centering
  \includegraphics[width=0.8\linewidth]{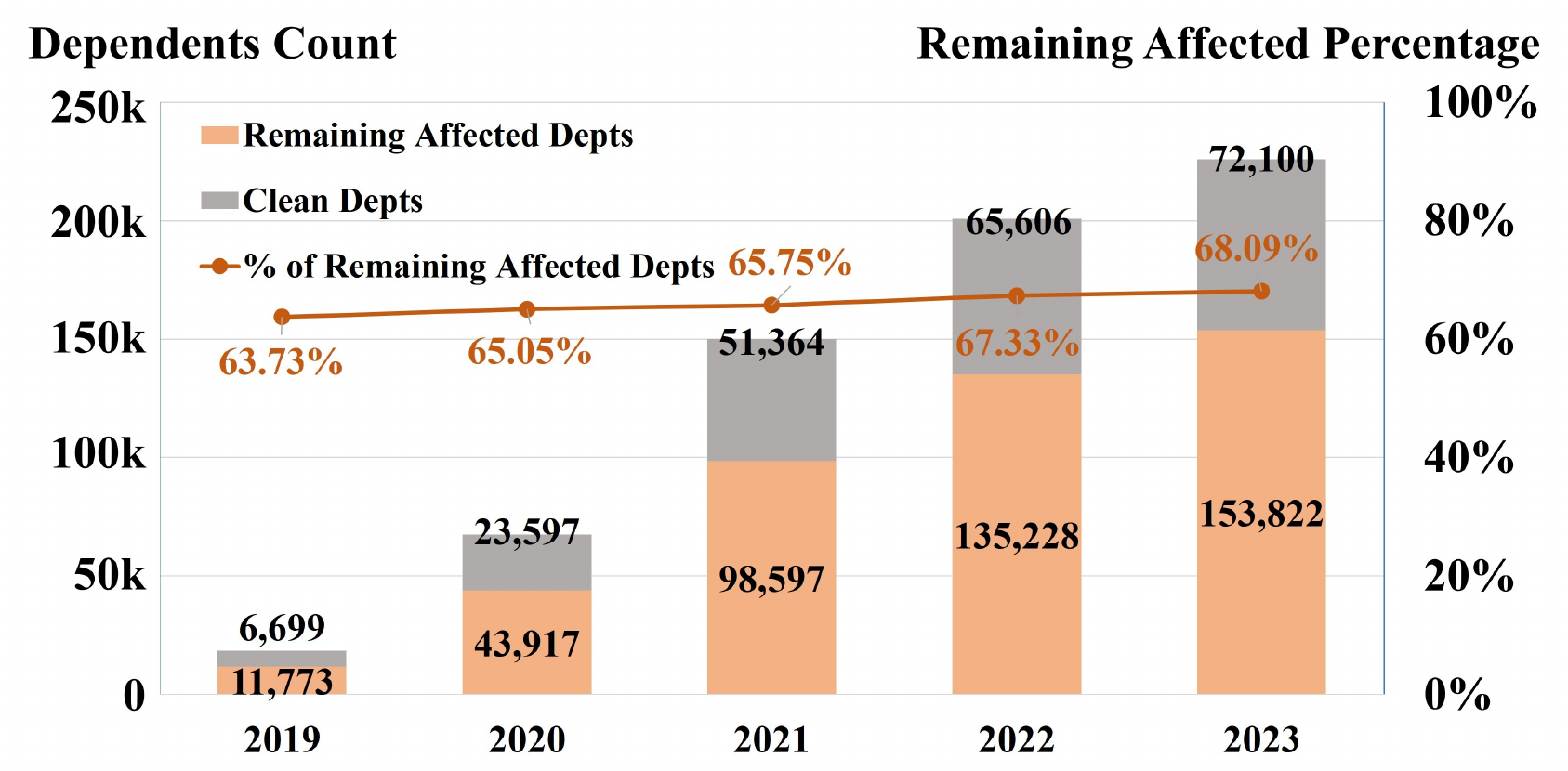}
  \caption{Impact over Time of Vulnerabilities from 2019}
  \label{fig:2019_vul}
\end{figure}

\chengwei{To demonstrate the impact of vulnerabilities occurring after 2019, we depicted the distribution of affected dependents in Figure~\ref{fig:2019_vul}. 
Specifically, affected dependents modules that have vulnerable dependencies, and clean dependents are on the opposite.}
It is seen that both affected and clean dependents increased substantially over time. Even if some of the vulnerable dependents got fixed over time, more dependents were developed and involved with the development of the Golang ecosystem. Hence, the proportions of affected dependents maintained at a steady level, but the number of affected dependents increased remarkably. It is proven that the legacy vulnerabilities have been persistently affecting increasing dependents in the ecosystem.

It is seen that both affected and clean dependents increased substantially over time. Even if some of the vulnerable dependents got fixed over time, \chengwei{more modules that depend on these vulnerable modules were introduced into the Golang ecosystem.} Hence, the proportions of affected dependents maintained at a steady level, but the number of affected dependents increased remarkably. It is proven that the legacy vulnerabilities have been persistently affecting increasing dependents in the ecosystem.

\begin{tcolorbox}[size=title,opacityfill=0.1,breakable,boxsep=1mm]
{\bfseries{Finding-2:}} 
Even for vulnerabilities of 2019 and before, the affected dependents have been increasing over the years while keeping the proportion of affected vulnerabilities steady.
\end{tcolorbox}

\subsection{RQ2: Patch Lagging Analysis}
Due to the significant amount of unresolved vulnerabilities in downstream dependencies over a long period, we further investigated the lags from the perspective of module maintainers. In this section, we employed the timestamps $T_{fix}$ (fixing commit time), $T_{ver}$ (version release time), and $T_{index}$ (Golang Index publish time) to calculate the lags $Lag_{ver}$ and $Lag_{index}$.
\subsubsection{\textbf{Data Preparation}}
Besides the vulnerability data from RQ1 Section~\ref{sec:rq1}, to compute the lags between $T_{fix}$ and $T_{index}$, the fixing commits with timestamps were derived from the associated repositories and the publish time of the patched versions was crawled from Golang Index~\cite{golangindex}. Specifically, we successfully identified the fixing commits for $1,269$ vulnerabilities. Besides the fixing commits,  we searched for additional commits that shared the same commit message. These newly identified commits were then treated as supplementary fixing commits. Among these commits, the earliest fixing commit release time was denoted as $T_{fix}$
Considering that the Golang Index was launched on April 10, 2019, we excluded $96$ vulnerabilities that were fixed before this date, as the calculation of $Lag_{index}$ is not applicable for them after the launching date. After exclusion, we obtained a dataset of $1,014$ vulnerabilities for further analysis. These vulnerabilities were distributed across $387$ repositories. Notably, $85.2\%$ of them were assigned CVEs.

\subsubsection{\textbf{Analysis of $Lag_{ver}$}}
\begin{figure}[]
  \centering
  \includegraphics[width=0.9\linewidth]{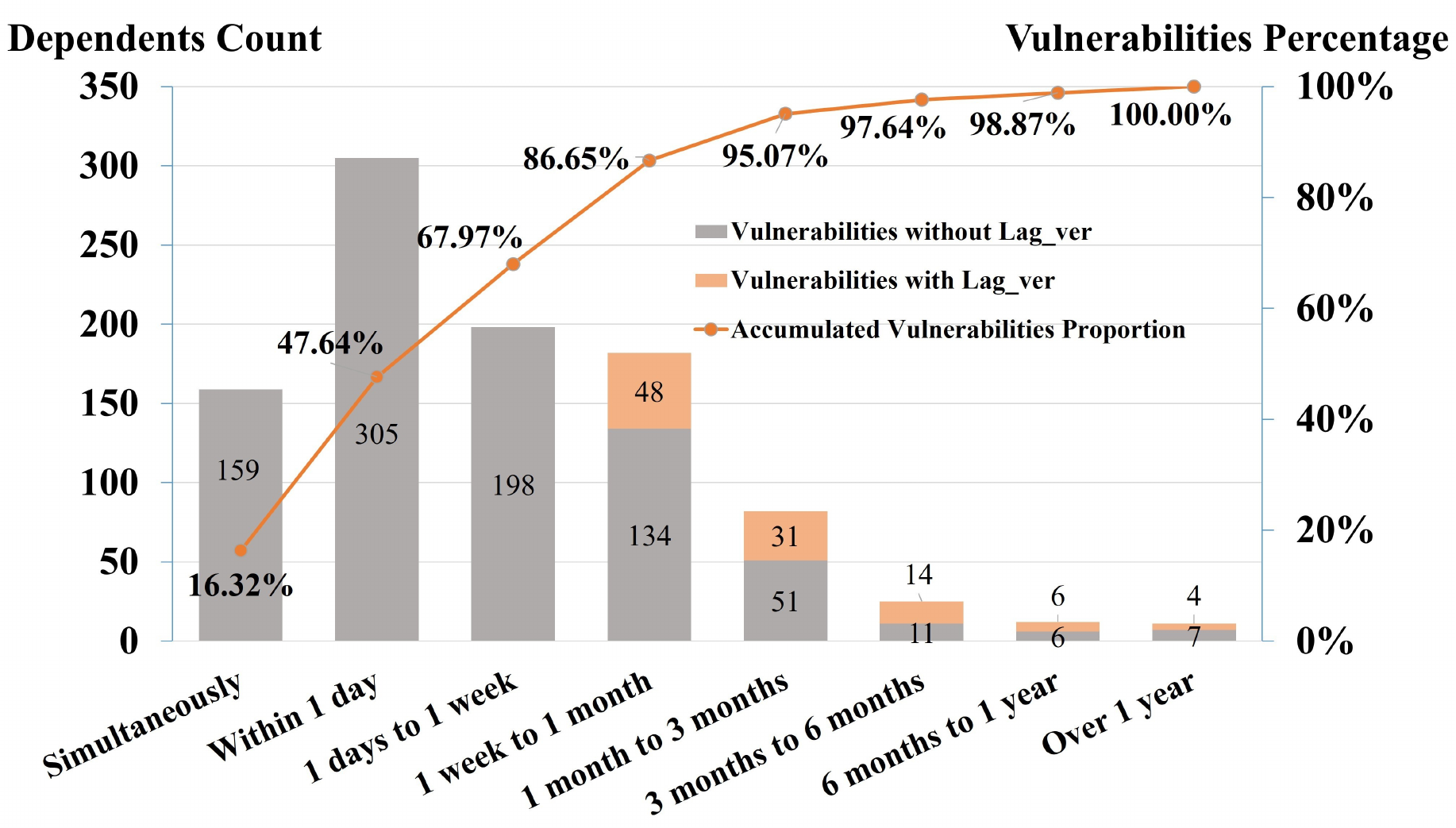}
  \caption{\chengwei{Distribution of $LT_{ver}$ and $Lag_{ver}$}}
  \label{fig:lag_ver}
  \small
  \begin{enumerate}
      \item \chengwei{Note that we only calculate $Lag_{ver}$ when Lead Time exceeds one week and the current cycle is greater than the normal cycle, i.e., the yellow bars.}
  \end{enumerate}
\end{figure}

We observed that modules of $40$ vulnerabilities have not released the patch version so that $Lag_{ver}$ could not be calculated. In Section~\ref{sec:rq4} (RQ4), we will clarify the reasons why no version was released. Among the remaining $974$ vulnerabilities that had at least one released patch version, we plotted the distribution of them based on the $LT_{ver}$ (time interval between the fixing commit and the release of the patch version) and $Lag_{ver}$ (Lead Time exceeds one week and the current cycle is greater than the normal cycle) in Figure~\ref{fig:lag_ver}.
To demonstrate the timeliness of version releases upon fixing commits, we roughly split the vulnerabilities into two segments, the $LT_{ver}$ within one week and beyond one week, as the testing and Continuous Integration checks could take time. We assume versions with $LT_{ver}<1week$ are timely patch releases for downstream users. 
Out of the $974$ vulnerabilities analyzed, $32.0\%$ of vulnerabilities did not have their patch versions released in a timely manner. It is noteworthy that $11$ vulnerabilities release patch versions over $1$ year after the fixing commit submission. 

\chengwei{Moreover, the distribution of $LT_{ver}$ by time intervals is presented in Figure~\ref{fig:lag_ver}. Notably, almost $47.64\%$ of vulnerabilities are fixed and tagged within 1 day, and $16.32\%$ of these vulnerabilities were even fixed and tagged simultaneously, which is pretty in time for downstream users. While there are also $130$ vulnerabilities had their patch versions released over $1$ month after the fixing commits, i.e., $LT_{ver} > 1 month$.
Because Golang would not automatically use the fixing commits as versions for dependencies to incorporate patches unless users explicitly declare them, the delayed version releases could result in the lag of fixing within the Golang ecosystem.}

\begin{tcolorbox}[size=title,opacityfill=0.1,breakable,boxsep=1mm]
{\bfseries{Finding-3:}}
\chengwei{$130$ vulnerabilities released patch versions in over 1 month after fixing commits, which could impede patch propagation in the Golang Ecosystem.}
\end{tcolorbox}

\chengwei{To unveil the reasons for delayed patch version releases, we assessed $Lag_{ver}$ within these vulnerable modules. After excluding vulnerabilities whose modules have no version released before fixing commit, $954$ vulnerabilities were retained for analysis. Among them, $734$ vulnerabilities had patch versions released within the expected cycles, while the rest $220$ had unusually long $LT_{ver}$. Notably, $103$ of them are with $LT_{ver} > 1week$, accounting for $10.8\%$ of all analyzed vulnerabilities, and their $Lag_{ver}$ as illustrated in Figure~\ref{fig:lag_ver}. We manually scrutinized the $62$ repositories that these vulnerabilities belong to, and observed that $58$ modules are still actively maintained in 2023, indicating that the delayed patch version releases have little correlation with the activeness of repositories.}


\chengwei{Moreover, because it is common that the fixing commits (i.e., pseudo-versions) are directly used as a temporary patch versions, we further verified if pseudo-versions are prevalent in these $62$ vulnerable modules so that it is not urgent to release patch versions. 
\chengwei{Out of $49$ vulnerable modules that are ever been imported by other projects,}
only $23$ repositories had dependents that used pseudo-versions to address vulnerabilities, which accounted for less than half of the vulnerabilities. This suggests that there could be lack of countermeasures for downstream users to patch vulnerable dependencies if the patch version tags are delayed.}


\begin{tcolorbox}[size=title,opacityfill=0.1,breakable,boxsep=1mm]
{\bfseries{Finding-4:}} 
\chengwei{Modules of $10.8\%$ of vulnerabilities had obvious patch version release lags, while only less than half of the corresponding modules are proactively updated by dependents via pseudo versions, posing a lack of countermeasures to promote the distribution of patches.}
\end{tcolorbox}

\subsubsection{\textbf{Analysis of $Lag_{index}$}}
\label{sec:rq2_index}
From the $1,014$ vulnerabilities. We first excluded $66$ vulnerabilities which had negative $Lag_{index}$ due to the version re-tagging after indexing.  
Out of the remaining $948$ vulnerabilities, $67.09\%$ of them have the $Lag_{index}$ within a week, 
\jc{indicating that the patch versions were published in the Golang Index promptly.}
Specifically, $297$ ($31.33\%$) of them have $Lag_{index}$ within one hour. \chengwei{We manually went through repositories of these vulnerabilities and found that 135 out of 297 within 65 repositories have integrated continuous integration. The rest was highly likely to be performed by maintainers themselves to declare the versions in \texttt{go.mod} or use \texttt{go~get} to index versions right after the release.} This proactive approach by maintainers ensured the registration of the patch version in the Golang Index, making it promptly available to downstream users. 


\begin{figure}[]
  \centering
  \includegraphics[width=0.9\linewidth]{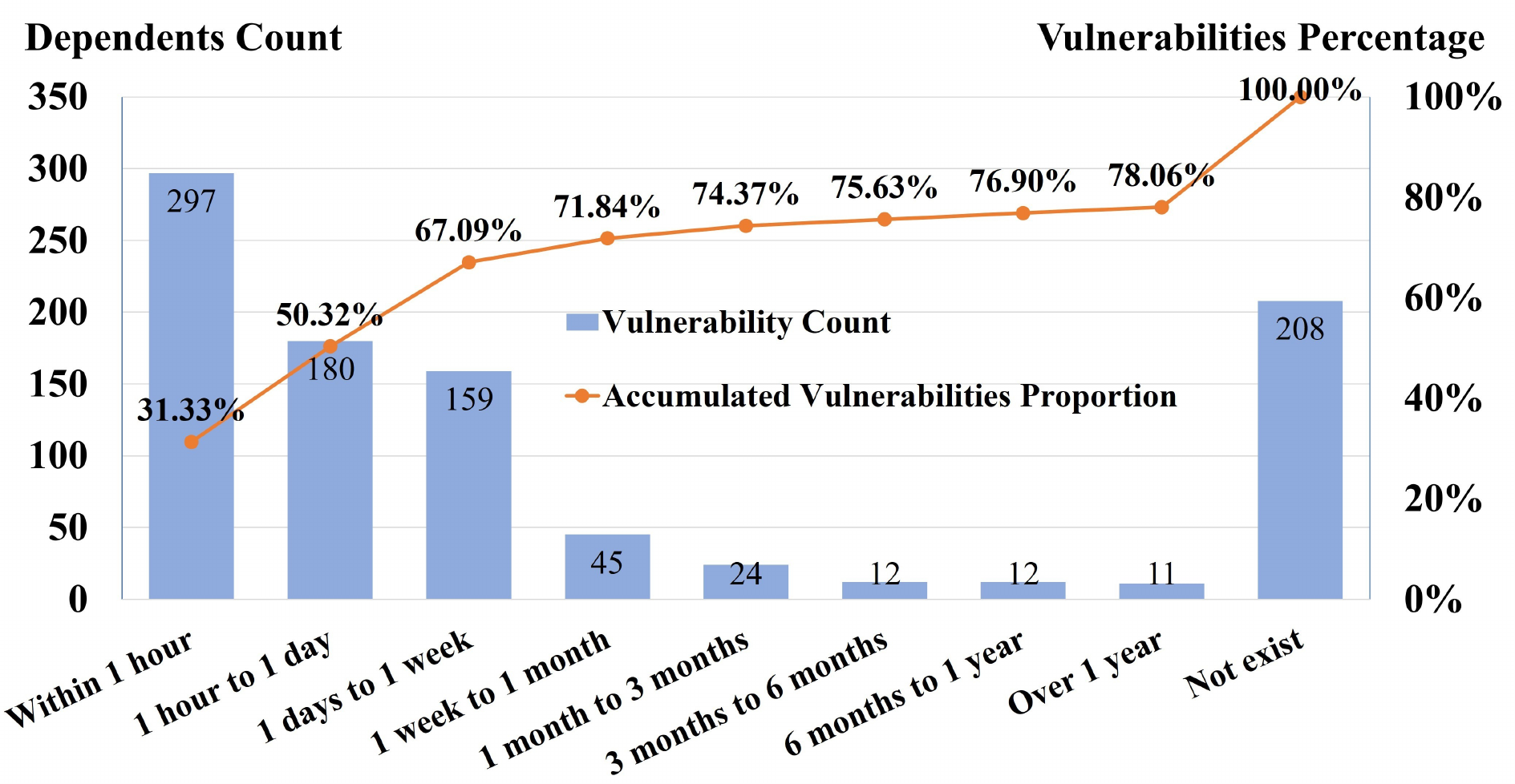}
  \caption{Distribution of $Lag_{index}$}
  \label{fig:lag_index}
\end{figure}
On the contrary, there were $208$ ($21.94\%$) of vulnerabilities whose $T_{index}$ was not available. These vulnerabilities had patch versions that were never utilized by users, and consequently, were not recorded in the Golang Index. As a result, users were unable to automatically access these patch versions via the Golang commands, hindering their ability to effectively address the associated vulnerabilities. 
\chengwei{Considering that the patch versions are only indexed upon usage,} 
we surmise that these modules could have limited users. We then collected their dependents to understand whether and how they addressed the vulnerabilities. It turned out that modules of $127$ of these vulnerabilities had a total of $136,480$ dependents, \chengwei{and surprisingly, only $136$ dependents have fixed the vulnerabilities via updating pseudo versions.}
Hence, the unindexed patch versions were mostly caused by rare usage and absent remediation.

\begin{tcolorbox}[size=title,opacityfill=0.1,breakable,boxsep=1mm]
{\bfseries{Finding-5:}} A majority ($67.09\%$) of vulnerabilities, demonstrated a swift patch version indexing. However, it is concerning that the patch versions of $21.94\%$ of vulnerabilities were not indexed in the Golang Index.
\end{tcolorbox}

\subsection{RQ3: Dependents Vulnerability Fixing}
\label{sec:rq3}
Apart from the perspective of maintainers, the adoption of patch versions by downstream users also greatly determines the life cycles of vulnerabilities. In this section, we conducted an analysis of how users address the vulnerabilities at a finer granularity by investigating the modification history of the Golang dependency manifest file \texttt{go.sum} for each affected dependent module. We first summarized the methods employed by users and calculated the proportions of them over all affected dependents.

\subsubsection{\textbf{Data Preparation}}
To begin our analysis, we gathered and downloaded all dependents stored in GitHub that met the criteria of having a vulnerability with a $T_{fix}$ date after April 10, 2019. This entailed downloading a total of $451,013$ dependents and $411,860$ succeeded for further analysis and investigation. Out of them, $253,865$ dependents had both \texttt{go.mod} and \texttt{go.sum} files, indicating that they were using Go Modules to manage their dependencies. However, for other dependents that lacked the \textit{go.sum} file, we were unable to obtain the modification history of the dependency relations. Therefore, we excluded these dependents from our analysis. We applied Algorithm \ref{algorithm:Lags} to determine $T_{dept}$, timestamp of the latest commit that changes \texttt{go.sum} for each dependent. 


\subsubsection{\textbf{Analysis of Fixes by Dependents}}

Generally, it is worth noting that only $11,446$ ($4.51\%$) successfully resolved all vulnerabilities we had collected. However, a significant number of dependents, $182,461$, ($71.87\%$) had not addressed any of the vulnerabilities.

\begin{figure}[]
  \centering
  \includegraphics[width=0.8\linewidth]{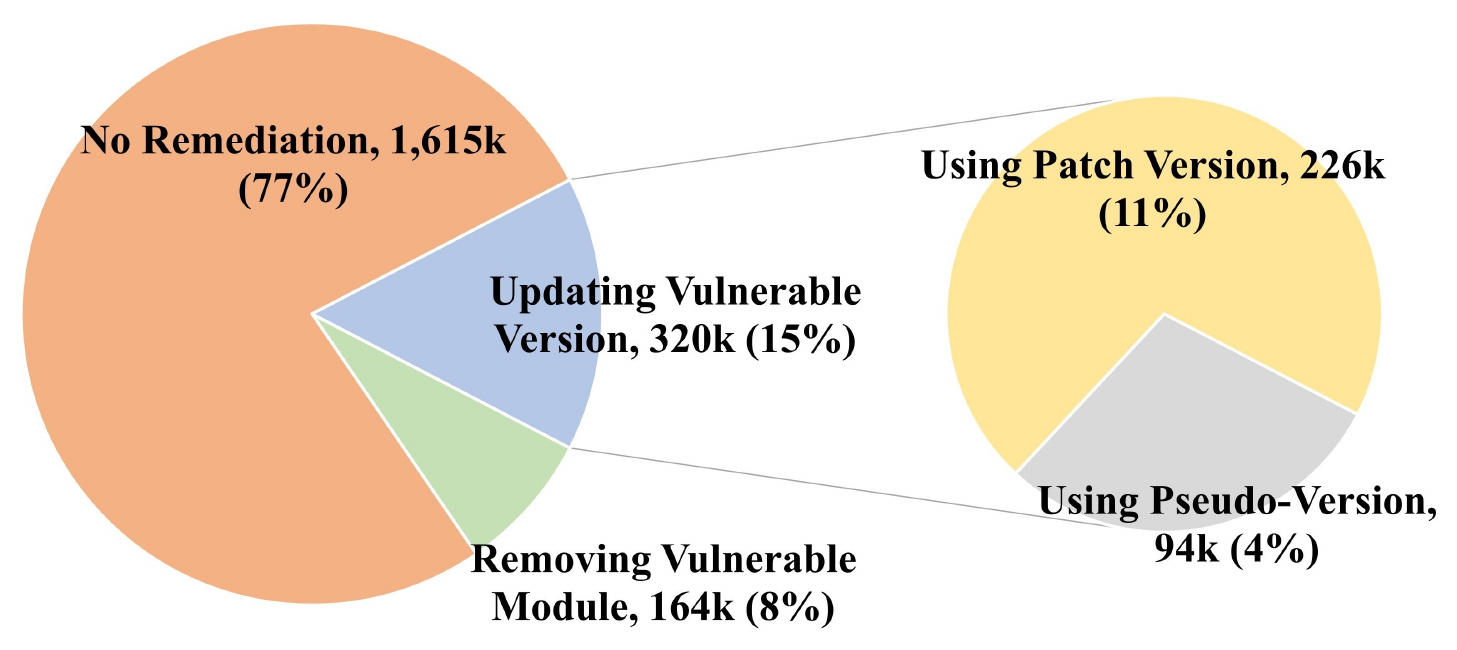}
  \caption{Distribution of Vulnerability-Dependent}
  \label{fig:rq3-fig1}
\end{figure}

In order to account for the number of dependents each vulnerability has, we quantified the vulnerability and dependent mappings using a unit that amalgamates a vulnerability with its dependent (referred to as a vulnerability-dependent, or VD). This approach resulted in a total of $744$ vulnerabilities being included in our further examination.
As illustrated in Figure~\ref{fig:rq3-fig1}, our analysis revealed that $163,930$ VDs ($7.81\%$) addressed the vulnerability by removing the vulnerable module from the \texttt{go.sum} file. Additionally, $319,961$ VDs ($15.24\%$) resolved the vulnerability by updating the vulnerable version. 
Among these VDs, 93,823 ($29.32\%$) used the pseudo-version mechanism, while 226,138 ($70.68\%$) used the patch version. Interestingly, the majority of VDs ($76.95\%$) still retained the vulnerability without taking any measures.

\begin{tcolorbox}[size=title,opacityfill=0.1,breakable,boxsep=1mm]
{\bfseries{Finding-6:}} 
Only nearly a quarter of the VDs have implemented measures to address the vulnerabilities. 
Among these VDs, 
the majority ($66.12\%$) opted to update the vulnerable version. Out of them, $70.68\%$ utilized the patch version to fix the vulnerabilities instead of utilizing the pseudo-version.
\end{tcolorbox}

\begin{figure}[]
  \centering
  \includegraphics[width=0.9\linewidth]{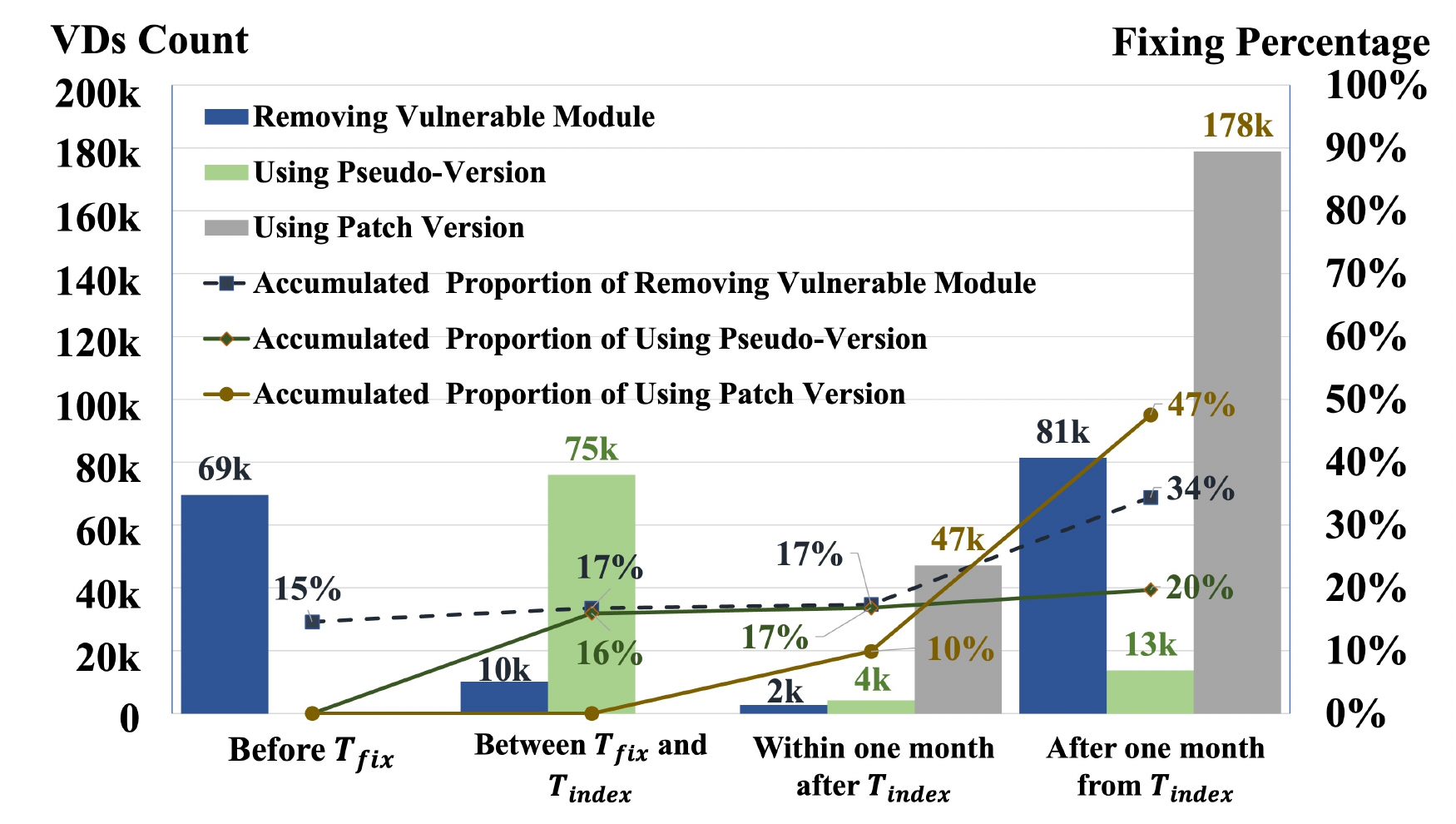}
  \caption{\chengwei{Distribution of Vulnerability-Dependents against $T_{dept}$}}
  \label{fig:rq3-fig2}
\end{figure}

\jc{To assess the effectiveness of fixing commits and indexed patch versions, we tracked changes in the number of dependents over time. }
Using algorithm~\ref{algorithm:Lags}, we obtained the $T_{dept}$ for each VD. 
\jc{To highlight the variations over time,}
we divided the timeline 
\jc{into 4 periods related to $T_{dept}$: \textit{Before $T_{fix}$}, \textit{Between $T_{fix}$ and $T_{index}$}, \textit{Within one month after $T_{index}$}, and \textit{After one month from $T_{index}$}, as depicted in Figure~\ref{fig:rq3-fig2}.}
Note that $127$ out of the $744$ vulnerabilities did not have patch version indexed thus no $T_{index}$ was available. Thus, they were categorized in \textit{Between $T_{fix}$ and $T_{index}$}.



\chengwei{When $T_{dept}$ is earlier than $T_{fix}$, removing the vulnerable modules is the only method to address the vulnerabilities. Our observations indicate that $14.61\%$ of the total $483,891$ VDs employed this approach during the specified period. Overall, we found that $34.42\%$ of VDs addressed vulnerabilities by removing the vulnerable modules, making the usage even out over the timeline. We speculate the reasons that removing modules could be caused by debloating or other maintenance considerations so that removing modules has an insignificant relationship with fixing vulnerabilities.}

In cases where $T_{dept}$ falls between $T_{fix}$ and $T_{index}$, the only available option to address the vulnerabilities while still preserving the modules was to upgrade to a pseudo-version. In this interval, pseudo-versions were employed to rectify vulnerabilities for $15.95\%$ of VDs. It's noteworthy that according to the Golang documentation~\cite{GoModule83:online}, a pseudo-version does not constitute a stable release. In totality, a mere $19.70\%$ of VDs opted for this method, the majority of which occurred during the period between $T_{fix}$ and $T_{index}$.

When the indexed patch version is out, $T_{dept}$ falls behind $T_{index}$. It is noteworthy that $9.92\%$ of VDs upgraded vulnerable modules to patch versions within the first month, but there is a boost of patch version adoption after the first month by $47.49\%$. 
This indicates that when patch versions were released and made available in the Golang Index, users were more inclined to fix the vulnerabilities by updating vulnerable modules with patch versions. 

In conclusion, when the patch version is available, users are more likely to adopt the patch version for vulnerability fixing. While the patch version is unavailable, users prefer the pseudo-version over removing modules. Although Golang allows the pseudo-version to serve as the fix at the first available moment, users still prefer the patch version based on SemVer, which further emphasizes the need to release the patch versions as soon as the fixing commit is out. 


\begin{tcolorbox}[size=title,opacityfill=0.1,breakable,boxsep=1mm]
{\bfseries{Finding-7:}} Using the patch version was the preferred method for fixing vulnerabilities among most users, with $47.49\%$ of all VDs. Maintainers should release patch versions as soon as possible when fixing commits are available.

\end{tcolorbox}

\subsection{RQ4: Fixing Lagging Inquiry}
\label{sec:rq4}
In the preceding sections, we quantitatively highlighted the lags associated with vulnerability fixes and explore how both maintainers and dependents contributed to these lags. Specifically, in this section, \chengwei{we delve into the reasons for 
1) the absence of patch version releases, 2) the delays of patch version releases, 3) the missing indexing of patch version to Golang Index, and 4) the missing patches to vulnerable dependencies, by combining inquiries to maintainers and manual analysis. Note that, generally, in all inquirying issues, we follow the similar structure: First, we declared us as a research team (to keep anonymous), and presented the information of specific vulnerabilities, including the CVE id and reference links. Next, for each specific types of lags, we explain the corresponding lags and potential threats to downstream users. After that, we either provided our recommendations on countermeasures and ask for their opinion, or directly asking for the reasons for corresponding delays.}
Since the community has been continuously replying to our questions, we would update the latest feeds at our website~\cite{dataset}.


\subsubsection{\textbf{Absent Patch Version Releases}}
\label{sec:4.1}
\chengwei{In our analysis of the 40 vulnerabilities that were not released with a patch version, we proactively engaged with the maintainers by submitting issues to inquire about the reasons for the absence. Among these, modules of 20 vulnerabilities have never released any tag, indicating that all dependents have to use pseudo-versions. 
These modules were excluded because the $Lag_{ver}$ was not applicable.}


For the remaining vulnerabilities, we submitted 20 queries via GitHub issue and received responses from \chengwei{8 maintainers. In these issues, we \ding{172} initiated 
elaborate the situation that no patch versions were released for the vulnerabilities even if they were already fixed with commits,
\ding{173} such missing patch version could result in users being unable to conveniently retrieve the latest patch versions, i.e., via the Golang tool "go list", and \ding{174} we recommend that maintainers to tag patch versions for these fixing commits to facilitate downstream users fixing the vulnerabilities. 
The detailed query structures are available on our website~\cite{dataset}.}

Among these responses, developers of 3 repositories agreed and replied to us with the patch versions. Regarding the other \chengwei{5} queries, one maintainer agreed with our observations and proposed that users could use the branch name \textit{master} as the version to obtain the latest pseudo-versions. 
However, this is considered provisional as Go Module would automatically resolve \textit{master} as the latest pseudo-version within the branch and update the \texttt{go.mod} file with the pseudo version number without assurance of compatibility.
Thus the recommended course of action is for maintainers to release a dedicated patch version for downstream users. In contrast, one maintainer responded that the project is not designed to be used as a library for other users, and therefore, the maintainer does not prioritize addressing the vulnerability for external users. The rest \chengwei{3} maintainers closed the queries without providing any response.


\begin{tcolorbox}[size=title,opacityfill=0.1,breakable,boxsep=1mm]
{\bfseries{Finding-8:}} 
Considering that developers from 3 repositories have not previously disclosed their patch versions, it could facilitate the adoption of patches if maintainers explicitly mention the vulnerability fixing in patch releases.
\end{tcolorbox}

\subsubsection{\textbf{Delayed Patch Version Releases}}
\chengwei{Next, we investigated the reasons behind the delayed release of patch versions. Conservatively, all vulnerabilities with $LT_{ver}> 1$ month were included in this analysis. Specifically, we manually examined the release notes and commit branches of the patch versions, resulting in 103 vulnerabilities to further categorize the reasons.} 

\chengwei{Typically, there are two typical types of delays for tagging patch versions. 1) For vulnerabilities whose fixing commits are followed with commits that changes functional codes 
before version tagging commits, we suspect these delays are waiting for \ding{172} \textbf{Other commits have to be incorporated into one release}. 
2) For vulnerabilities whose fixing commits are not followed with commits of code-related changes till version tagging commits, there could be two types of reasons that are subjective to maintainers, \ding{173} \textbf{Issues with testing and CI checking}: the maintainers need more time to examine the changes, and \ding{174} \textbf{By convention, infrequent release of versions}: they are just waiting for regular releases. }

\chengwei{Therefore, we take them as the primary reasons for patch version tagging delays, 
and since patch versions are just delayed instead of absent, after elaborating on the delays of specific vulnerabilities, we directly ask the maintainers for their reasons by choosing from these primary reasons or give \ding{175} \textbf{Other} reasons, via GitHub issues. After excluding 7 vulnerabilities we are unable to reach maintainers (i.e., read-only repositories or no issues allowed), and 9 vulnerabilities are fixed before the first version tag (i.e., no release cycles), issues are submitted to the repositories of $87$ vulnerabilities.}

\chengwei{Till Nov 2023, we only received responses of reasons for 11 vulnerabilities. However, surprisingly, all these responses have proactively provided much more detailed reasons to us. 1) (3 cases) The patch versions we found delays are actually not tagged on the version branch where the vulnerabilities are first discovered and patched, and they also provided the first patch versions. However, based on our observation, it still takes a very long time for them to be merged into the main branches. 2) (2 cases) the vulnerabilities were first identified in a development branch for the next minor versions, and they had to wait for the first stable versions before tagging, however, they have quickly back-ported the patches to previous version tracks and the follow-up patch versions are quickly tagged.
3) (2 cases) These vulnerabilities are not already patched by their intended users, so no need to push the patch versions to the public in a fast pace. 4) (2 cases) These patch versions are lagged due to the lack of availability of maintenance.}

\chengwei{These reasons indicate the correlation between the complex repository management strategies and the delays of patch version tagging. Therefore, we conducted a manual analysis of the locations of fixing commits and patch versions for all patch version tagging, and further categorized these reasons. 
(1) Multiple version tracks, sometimes even the minor version tracks, are maintained on different branches, while the delays of patch version tagging among them varies (37 cases). 
(2) Vulnerabilities are fixed during the development of new versions, these patch commits are unable to be tagged before stable version tags (12 cases). 
(3) Fixing commits are committed in different branches, and it takes more time to backport patches to other branches (9 cases).
(4) Some patches are introduced when conducting breaking changes, such as upgrading to next major version (3 cases).
Apart from these, there are still 40 cases we are unable to identify explicit reasons, and we will continue to update the feedback from maintainers on our website~\cite{dataset}.
}

\begin{tcolorbox}[size=title,opacityfill=0.1,breakable,boxsep=1mm]
{\bfseries{Finding-9:}} 
\chengwei{The complex repository management strategies contribute the most to the delays of patch version tagging, especially the inconsistent version management on branches for different purposes.}

\end{tcolorbox}

\subsubsection{\textbf{Not Pushing Patch Version to Golang Index}}
As revealed in Section~\ref{sec:rq2_index}, we identified 208 vulnerabilities whose released patch versions were not indexed in Golang.
\chengwei{Since modules could be indexed only when they are imported by other projects, 
we excluded vulnerabilities whose patch versions are not in Golang Index, and analyze the usage of vulnerable versions of the corresponding modules of the rest 112 vulnerabilities from 30 repositories.}


\chengwei{Since maintainers could easily index the patch versions by \texttt{go get} command, it is critical to understand the reasons for this absent action and inform them of the criticality. Before submitting inquiries, we manually examined the repositories to seek possible reasons. 
We found 4 repositories (6 vulnerabilities) stated that they were applications, not third-party modules, implying no intent to index patch versions. Additionally, 6 repositories (7 vulnerabilities) used version tags that do not follow Golang versioning requirements~\cite{GoModule7:online}, rendering their patch versions unindexable.
}


\chengwei{After excluding these cases, we open issues for the remaining 99 vulnerabilities. Specifically, after stating our observations that their patch versions for specific vulnerabilities are not indexed in Golang Index, we advised maintainers to leverage Golang commands like \textit{go get module@version} or similar methods to facilitate patch version indexing after tagging patch version for vulnerabilities.}


\chengwei{Till Nov 2023, we received responses for 37 vulnerabilities. It is noteworthy that we get positive responses on the necessity of indexing patch versions. Specifically, responses on 10 vulnerabilities indicated their willing to take our suggestions to execute the \textit{go get} command, 3 of them have pushed patch versions to Golang Index and the rest 7 sent relative information to their security teams. The responses on the rest vulnerabilities did not take actions due to various reasons: 1) responses of 15 vulnerabilities replied that their projects are products instead of third-party modules, while according to our data, they have hundreds of dependents importing vulnerable versions; 2) responses of 7 vulnerabilities indicated that their modules are not directly imported via Go Module and they have already provided necessary countermeasures to facilitate downstream use. 3) responses of 5 vulnerabilities stated they are not the correct repositories to raise issues due to migration or vulnerability reference error. Given that all maintainers replied with positive feedback, 
it is anticipated that the timely indexing of patch versions should be welcomed by the community. However, considering that not all maintainers are willing to vulunteerly do so, there still lacks proper mechanisms to actively search and index such new patch versions to facilitate in-time distribution.} 




\begin{tcolorbox}[size=title,opacityfill=0.1,breakable,boxsep=1mm]
{\bfseries{Finding-10:}} All maintainers who responded agreed with our observations and suggestions. Nevertheless, to facilitate patch propagation, the Golang release mechanism should index new versions as soon as possible without relying on maintainers.
\end{tcolorbox}

\subsubsection{\textbf{Not Fixing the Vulnerability}}
\chengwei{To ensure the popularity, }we first selected the top 120 most-starred dependent repositories of vulnerable modules for analysis, \chengwei{and for each dependent, we localize the vulnerabilities in their dependencies based on our dependency relations.}
If vulnerable dependencies remained unaddressed by the data collection date, \chengwei{we notified maintainers by submitting issues detailing the persistence of vulnerable dependencies and the available patch versions with recommended dependency upgrades.} Considering that different vulnerabilities have different impact and exploitability, \chengwei{we made inquiries respectively based on VDs}. In total, 337 VDs (120 dependents) were queried. 


\chengwei{Among the 337 VDs analyzed, we received responses from maintainers of \chengwei{235} VDs (\chengwei{72} dependents). Specifically, maintainers of \chengwei{202} VDs (\chengwei{61} dependents) accepted suggestions to update the vulnerable dependencies. 
}
However, in the responses of 16 VDs (8 dependents), maintainers disagreed with update suggestions for the following reasons: 
(1) The projects were not using the vulnerable functions (\chengwei{10} VDs, \chengwei{4} dependents); 
(2) The vulnerable modules were rarely used, and addressing the vulnerabilities would require updating the Go version, which could be troublesome (4 VDs, 2 dependent); 
(3) There was a compatibility issue preventing upgrading the module (1 VD, 1 dependent); (4) The maintainer suggested downstream users address the vulnerability by themselves (1 VD, 1 dependent).

\begin{tcolorbox}[size=title,opacityfill=0.1,breakable,boxsep=1mm]
{\bfseries{Finding-11:}} The majority of maintainers (89.95\%) addressed the vulnerabilities in a timely manner once they became aware of the presence of vulnerabilities in their projects.
\end{tcolorbox}







\section{Discussion}

\subsection{Recommendations}
\label{recommendations}

\chengwei{As an early adopter of decentralized package management, Golang originally encouraged direct repository imports as dependencies by commit hashes (i.e., pseudo versions) without releases to public registries~\cite{godepmanage}. However, with the growing complexity of project size, Go Modules introduced the SemVer tags for better version control management~\cite{Semanticgo}, leading to a chaotic situation where both SemVer for compatibility concerns and pseudo-versions for swift development run on parallel as revealed in RQ1. Therefore, we highlight the recommended solutions for different stakeholders, as well as future research directions, to mitigate such severe situation.}


\noindent \chengwei{\textbf{Maintainer Perspective.} Several challenges arise due to the module distribution and versioning practices in the Golang ecosystem. Our study highlights significant delays in releasing vulnerability fixes (i.e., $Lag_{ver}$ and $Lag_{index}$), hindering the adoption of patch versions. To mitigate this, maintainers should prioritize timely tagging and indexing of updates. Additionally, 
as revealed in RQ4, some maintainers favor users importing directly from the \textit{master} branch to keep updated, which could result in compatibility issues. This calls a stronger focus to ensure compatibility.
Lastly, maintainers often delay tagging fixing commits until sufficient feature additions are made, risking user exposure to vulnerabilities. Balancing the urgency of vulnerability fixes with feature development in version releases is essential for maintaining ecosystem security.}

\noindent \chengwei{\textbf{User Perspective.} 
Vulnerabilities persist in many modules that once imported vulnerable dependencies, even though users often update to patch versions when available. This issue arises because \textit{go.sum} naturally locks the dependencies if they still satisfy the declarations in \textit{go.mod} to avoid incompatibility.
However, this can delay updates that include essential fixes. We suggest users periodically unlock their dependencies to allow timely updates. Additionally, our findings show that many users address vulnerabilities because they get alerts from other users or SCA tools like Dependabot. Therefore, we recommend incorporating vulnerability checks from third-party auditors or SCA tools during their testing, i.e., integrated in CI/CD pipelines, to ensure security updates are promptly applied. }

\noindent \chengwei{\textbf{OSS governance Perspective.} The absence of a centralized registry in Golang limits its ability to track vulnerabilities in modules. This is crucial given the increasing emphasis on supply chain security.} 
\chengwei{Other ecosystems, like Maven and NPM, have integrated security tools and databases, such as CVE mappings and \textit{npm audit}, enhancing their security posture. Go Module could benefit from similar features. Implementing a database for vulnerability and patch data would aid in distributing fixes to users. }
\chengwei{Moreover, addressing compatibility issues arising from using pseudo versions for fixing vulnerabilities could be achieved by incorporating compatibility-checking tools for dependency upgrades.}

\chengwei{
We also highlight some possible research directions.
1) Vulnerability Data Enhancement: 
In the data preparation, considerable vulnerabilities could not be associated with their partial or all fixing commits.
Vulnerability data quality would be critical to conduct finer-grained monitoring and governance for the ecosystems. Therefore, techniques for identifying these patches, or even the root cause (i.e., vulnerable functions) and proof of concepts (POCs), could be further researched.
2) In-depth Vulnerability Impact Analysis: Given the frequent reuse of vulnerable modules in Golang, it is essential to introduce more nuanced analysis, such as vulnerability reachability analysis. These would boost identifying and prioritizing the most critical vulnerabilities for timely remediation.
3) Robust Compatibility Checks: 
Compatibility concerns are paramount when updating pseudo versions in Golang. Implementing rigorous compatibility checks prior to upgrades can alleviate user apprehensions and encourage the adoption of patches. 
Furthermore, SCA tools should include pseudo-version upgrades as part of their vulnerability remediation toolkit, particularly when patch versions are not available and compatibility can be thoroughly verified.
}

\subsection{Limitations and Threats to Validity}

\chengwei{1) Since Golang Index only record modules that are ever imported as dependencies, we are unable to find Golang modules if they are not never used by others. To ensure data completeness, we retrieve modules and dependencies from Open Source Insight~\cite{osinsightdata}, which collects these data from various sources. Our dataset is also statistically more complete than \textit{libraries.io}, which further ensure the representativeness.
2) There could be vulnerabilities that are never discovered or reported, we are unable to perfectly find out all vulnerabilities. There are still considerable vulnerabilities that we are unable to locate their fixing commits based on existing information. Even so, we have tried our best to collected vulnerability information from the most mainstream platforms and validate the coverage by comparing with OSV and GitHub advisory. 
3) Downstream users could accidentally remove vulnerable dependencies for other purpose instead of excluding vulnerabilities, which could bias our analysis. However, this would not influence the conclusions since the majority of dependents are still vulnerable even if we take that as a fix.
4) Apart from the 66 cases whose version tag time is later than indexing time, there could be more cases of re-tagging on GitHub, which could underestimate the existence of $Lag_{index}$, while There is no practical solution to accurately check potential re-tagging in all repositories. Moreover, even so, our conclusion on the existence of $Lag_{index}$ would not be affected.
5) We introduced ourselves in the issues raised for maintainers by mentioning we were a research team on Golang, which might cause biases in the responses. We have manually processed the responses to attempt to eliminate the biases from the summarized results.}

\section{Related Work}
\noindent $\bullet$ \textbf{Vulnerability Analysis in Ecosystem.}
Numerous research studies
have delved into the security of open-source software ecosystems. 
\chengwei{
Decan et al.~\cite{decan2018impact} conducted an empirical study to unveil the severity of vulnerability propagation in the NPM ecosystem.}
Alfadel et al.~\cite{Alfadel2021} analyzed security vulnerabilities in the PyPI ecosystem and found over $50\%$ were patched after public disclosure. 
\chengwei{Zhang et al.~\cite{zhang2023mitigating, zhang2023compatible} analyzed the persistent vulnerabilities in the Maven ecosystem and proposed \textit{Ranger} to restore secure version ranges against the vulnerabilities.
Wu et al.~\cite{wu2023understanding} studied the reachability of Maven vulnerabilities and found 73\% of vulnerabilities are not reachable and safe for downstream users.
Reid et al.~\cite{reid2022extent} leveraged a file-level code reuse detector to locate the extensive vulnerable reused code in software projects in multiple languages, which severely jeopardizes the security of OSS. 
Fitzgerald et al.~\cite{fitzgerald2019towards} proposed a knowledge flow graph that connects libraries, files, projects, authors, and code instances together to measure the multi-facet Free/libre open source ecosystem.}
Tan et al.\cite{Tan2022} found that over $80\%$ of affected CVE-Branch pairs remained unpatched in OSS projects. 
Xu et al.\cite{Xu2022} studied CLV issues in PyPI and Maven ecosystems, identifying $82,951$ projects dependent on vulnerable C project versions. 
Shahzad et al.\cite{Shahzad2020} measured a large vulnerability dataset, showing the prevalence of DoS and EXE vulnerabilities and an increase in SQL, XSS, and PHP vulnerabilities. 
\chengwei{
Existing studies mainly examine ecosystem security by either reasoning on dependency networks constructed from centralized package managers or identifying the distribution of vulnerable code across artifacts.
Instead, Golang, as the first ecosystem managed by decentralized registries while integrated with centralized indexing, its unique vulnerability life cycles and management strategies are firstly studied in this paper.}


\noindent $\bullet$ \textbf{Technical Lag Studies.}
attempted to compare the time of the CVE publish, security fix and publication of a release that includes the fix. 
Imtiaz et al.~\cite{Imtiaz2022} studied technical lag, including time between fix and release, documentation in release notes, code change characteristics, and time between release and advisory publication. They found that the median security release becomes available within $4$ days, $61.5\%$ of security releases come with release notes documenting security fixes, and $13.2\%$ indicated backward incompatibility through semantic versioning, with $6.4\%$ mentioning breaking changes in the release notes.
Chinthanet et al.~\cite{Chinthanet2021} investigated lags between vulnerable and fixing releases. They found that fixing releases are rarely standalone, with up to $85.72\%$ of bundled commits unrelated to the fix. Stale clients require additional migration effort, even with quick package-side fixing releases.
Decan et al.~\cite{Decan2018} empirically studied technical lag in the npm dependency network, finding that package releases suffer from technical lag without benefiting from the latest dependency updates.
Previous studies ignored Golang's unique characteristics, where versions must be utilized before appearing in the Golang Index, setting it apart from other languages in versioning systems.

\noindent $\bullet$ \textbf{OSS Ecosystem-Related Studies.}
Researchers
have conducted in-depth studies on various types of ecosystems. 
Wang et al.\cite{Wang2021} developed HERO to investigate dependency management in Golang, achieving a high $98.5\%$ detection rate on DM issues. Cogo et al.\cite{Cogo2021} studied same-day releases in npm, finding important changes despite the restricted time frame, with some releases being error-prone due to large quick changes. Kula et al.\cite{Kula2018} explored developers' library dependency updates, discovering that $69\%$ of interviewees were unaware of vulnerable dependencies. Zhao et al.\cite{zhao2023software} studied the variety of Maven rule implementations in popular Java projects. 
Tang et al.~\cite{10.1145/3551349.3560432} systematically studied the third-party reuse of C/C++ libraries in different formats.
Wu et al.\cite{wu2023ossfp} have emphasized the fingerprint functions in software component detection for the C ecosystem. Zimmermann et al. Li et al.~\cite{li2023demystifying} studied the stable features of component reuse in the Rust ecosystem.
Sun et al.~\cite{10.1145/3611643.3616270} investigated the code reuse in the form of subcontract in Solidity.
\cite{Zimmermann} studied the potential impact of individual packages on the ecosystem due to vulnerable or malicious code in third-party dependencies.
Previous studies neglected Golang's unique versioning system, causing delayed patch version adoption and prolonging vulnerability life cycles.


\section{Conclusion}
Our analysis revealed that vulnerabilities significantly affected $66.10\%$ of the modules, while $62.85\%$ of the dependents had not addressed these vulnerabilities. We quantitatively proved that the timely patch release and indexing could greatly facilitate the patch adoption by downstream users. Through the inquiries about reasons behind lagged patch release, indexing, and adoption, \chengwei{we identified the deficiencies of countermeasures to fulfill the capability to accelerate the propagation of vulnerability patches in the Golang ecosystem. 
Practical recommendations for the perspectives of different stakeholders and possible research directions, are also concluded to facilitate the enhancement of security for the Golang ecosystem.}

\section*{Acknowledgments}
This study is supported under the RIE2020 Industry Alignment Fund – Industry Collaboration Projects (IAF-ICP) Funding Initiative, as well as cash and in-kind contribution from the industry partner(s). it is also supported by the National Research Foundation, Singapore, and DSO National Laboratories under the AI Singapore Programme (AISG Award No: AISG2-GC-2023-008). It is also supported by the National Research Foundation, Singapore, and the Cyber Security Agency under its National Cybersecurity R\&D Programme (NCRP25-P04-TAICeN) and the NRF Investigatorship NRF-NRFI06-2020-0001. Any opinions, findings and conclusions or recommendations expressed in this material are those of the author(s) and do not reflect the views of National Research Foundation, Singapore and Cyber Security Agency of Singapore.
\section*{Data Availability} 
The dataset and scripts can be accessed at our website~\cite{dataset}.

\bibliographystyle{ACM-Reference-Format}
 \bibliography{reference}
\end{document}